\documentstyle[12pt,aasms4]{article}

\def\be{\begin{equation}}
\def\ee{\end{equation}}
\def\ba{\begin{eqnarray}}
\def\ea{\end{eqnarray}}

\def\la{\mathrel{\mathpalette\fun <}}
\def\ga{\mathrel{\mathpalette\fun >}}
\def\fun#1#2{\lower3.6pt\vbox{\baselineskip0pt\lineskip.9pt
        \ialign{$\mathsurround=0pt#1\hfill##\hfil$\crcr#2\crcr\sim\crcr}}}

\slugcomment{To be submitted to ApJ}

\begin{document}
\null\vspace{-62pt}
\begin{flushright}
May 7, 1998; astro-ph/9804195
\end{flushright}

\title{A Catalog of Color-based Redshift Estimates \\
for $z \la 4$ Galaxies in the Hubble Deep Field}

\author{Yun Wang, Neta Bahcall, \& Edwin L. Turner}
\affil{{\it Princeton University Observatory} \\
{\it Peyton Hall, Princeton, NJ 08544\\}
{\it email: ywang,neta,elt@astro.princeton.edu}}

\vspace{.4in}
\centerline{\bf Abstract}
\begin{quotation}

We derive simple empirical color-redshift relations for $z\la 4$
galaxies in the Hubble Deep Field (HDF) using a linear function of
three photometric colors ($U-B$, $B-V$, $V-I$).
The dispersion between the estimated redshifts and the
spectroscopically observed ones is small for relations
derived in several separate color regimes; the dispersions
range from $\sigma_z\simeq 0.03$ to 0.1 for $z\la 2$ galaxies,
and from $\sigma_z\simeq 0.14$ to 0.25 for $z\ga 2 $ galaxies.
We apply the color-redshift relations to the HDF photometric catalog
and obtain estimated redshifts that are consistent with those
derived from spectral template fitting methods.
The advantage of these color-redshift relations
is that they are simple and easy to use and do not depend on
the assumption of any particular spectral templates; 
they provide model independent redshift estimates for 
$z\la 4$ galaxies using only multi-band photometry,
and they apply to about 90\% of all galaxies.
We provide a color-based estimated redshift catalog of HDF galaxies
to $z\la 4$.
We use the estimated redshifts
to investigate the redshift distribution of galaxies in the HDF;
we find peaks in the redshift distribution that suggest large-scale
clustering of galaxies to at least $z\sim 1$ and that are consistent
with those identified in spectroscopic probes of the HDF.

\end{quotation}


\section{Introduction}

The Hubble Deep Field (HDF)  provides accurate multi-band
photometry of galaxies to very faint magnitudes, with 
10$\sigma$ $AB$ magnitude limit of $m_{814}=27.60$ (\cite{HDF}).
The faint limit of the HDF makes it difficult
to obtain spectroscopic redshifts for the majority of the 
galaxies in the field.
It is therefore useful to derive estimated
redshifts of these galaxies using the available multi-band photometry.
Several groups (\cite{Lanze96}, \cite{GwHa96}, \cite{Saw97})
obtained photometric redshifts for the HDF 
by comparing the observed $UBVI$ fluxes of each object with 
a set of galaxy spectral templates of different galaxy types 
redshifted to evenly spaced redshifts.
Since spectroscopic redshifts have been
measured and published for $\sim$ 100 galaxies in the HDF
(\cite{Cohen96}, \cite{Hogg98}; \cite{Stei}; \cite{DEEP}),
it is possible to fit analytic expressions for photometric redshifts.
In this paper, we explore a simple empirical
approach to estimating redshifts of galaxies based on their colors
(see \cite{Conno97}, \cite{Brun97}, \cite{Conno95} for an alternative
empirical approach);
this method has the advantage of being simple, model independent
(i.e., it does not depend on the assumption of any particular
set of galaxy spectral templates),
and easy to use in determining approximate redshifts
of $z\la 4$ galaxies.

We determine the empirical analytic relations for color redshifts in \S 2.
We compare our estimated redshifts for the HDF galaxies
with those obtained by the template-fitting method in \S 3. We 
describe our estimated redshift catalog of HDF galaxies to
$z\la 4$ in \S 4 (the Web site address of the catalog is given).
We investigate the redshift clustering of HDF galaxies in \S 5,
and summarize our results in \S 6.

\section{Empirical Color-redshift Relations}

The HDF covers a 4 square arcmin area of sky in the northern
continuous viewing zone (\cite{HDF}).
The HDF galaxies have been observed in four passbands:
$m_{300}$ (which we denote by $U$), $m_{450}$ ($B$),
$m_{606}$ ($V$), and $m_{814}$ ($I$).
We use the HDF photometric catalog by Sawicki et al. (1997); 
it contains 848 galaxies with measured fluxes in all
four passbands to a magnitude limit of $I=27$.
Note that we use the AB system following Sawicki et al. (1997).

In deriving the color-based estimated redshift relations for the HDF,
we use 82 galaxies with measured spectroscopic redshifts in this
field, with redshifts in the
range $z\simeq 0.1$-3.5 (\cite{Cohen96}, \cite{Hogg98}; 
\cite{Stei}; \cite{DEEP}).
Five of the galaxies (with $2.8 \leq z \leq 3.5$) 
are used with only upper limits to their U fluxes (as derived
by Sawicki et. al. 1997).

We first divide the galaxy sample into regions of high and low redshifts
($z\ga 2$ and $z< 2$) based on empirical color cuts as follows. For
$z\ga 2$, the galaxies satisfy one of the following three color selection
criteria:
\ba
\label{eq:highzselect,1}
&& U\geq 25.66, \, U-B\geq 0.91, \, B-V\leq 1.37, \, V-I\leq 0.5;\\
\label{eq:highzselect,2}
&& I>23.5, \, U-B>2.2; \\
\label{eq:highzselect,3}
&& I>23.5, \, B-V>2.2, \, U-B>-0.5 .
\ea
Lower redshift galaxies, $z<2$, generally fall outside
these color-magnitude regions.
Note that the above color selection criteria for $z\ga 2$ galaxies
reflect our current knowledge from galaxies with measured spectroscopic
redshifts; they may need to be revised as new data becomes available.
To minimize the redshift dispersion in the empirical color-redshift
analytic fits, we further divide the regions into
two color ranges for $z\ga 2$ and three color ranges for $z< 2$.
These empirical divisions reflect the color shifts as a function
of redshift for different spectral type galaxies.
For each color range, we fit an analytic relation for the 
redshift, $z_a$, which is linear in color, 
\be
\label{eq:fit}
z_a = c_1 +c_2\,(U-B) +c_3\,(B-V) +c_4\,(V-I),
\ee
where $c_i$ ($i=1,4$) are constants. The error in $z_a$ due to photometric
errors is
\be
\Delta z_a = \sqrt{ (c_2\,\Delta U)^2+ \left[(c_3-c_2)\Delta B\right]^2
+\left[(c_4-c_3)\Delta V\right]^2 + (c_4\,\Delta I)^2 }.
\ee

We use Eq.(\ref{eq:fit}) to determine the best fit between 
the observed spectroscopic redshifts
of 82 HDF galaxies at $z\simeq 0.1 $-3.5 and their colors,
in each of the five separate color ranges. 
We find the following
best-fit relations and their redshift dispersions, $\sigma_z$.
The redshift dispersions are calculated using the jack knife method
(see \cite{Lupton} for a simple description),
which is also used to estimate the uncertainties of the coefficients
in the best-fit relations.
For easy reference, we assign each color range a number, $cr$,
$1\leq cr \leq 5$.

For the $z<2$ galaxies (see photometric range discussed above):
\newcounter{bean}
\setcounter{bean}{0}
\begin{list}%
{(\roman{bean})}{\usecounter{bean}}
\item{$cr$=1: $(U-B)<(B-V)-0.1$: \hskip 5cm (28 galaxies)
\be
\label{eq:Lz3,1}
z_a = 0.4111 -0.1852\,(U-B) -0.3062\,(B-V) +0.7301\,(V-I),\\
\hskip 1cm    
\sigma_z=  0.034.
\ee
The 1-$\sigma$ uncertainties of the coefficients are 
0.0036, 0.0058, 0.0084, and 0.0092 respectively.
}
\item{$cr$=2: $(U-B)\geq (B-V)-0.1>(V-I)$: \hskip 3cm (21 galaxies)
\be
\label{eq:Lz3,2}
z_a=   0.163   -0.171\,(U-B)   +0.340\,(B-V)  +0.194\,(V-I),\\
\hskip 1cm
\sigma_z=  0.095.
\ee
The 1-$\sigma$ uncertainties of the coefficients are 
0.013, 0.014, 0.045, and 0.047 respectively.
}
\item{$cr$=3: $(U-B)\geq (B-V)-0.1\leq (V-I)$: \hskip 3cm (19 galaxies)
\be
\label{eq:Lz3,3}
z_a=   1.126  +0.480 \,(U-B)  -0.513 \,(B-V) -0.250 \,(V-I),
\hskip 1cm
\sigma_z= 0.097.\\
\ee
The 1-$\sigma$ uncertainties of the coefficients are 
0.029, 0.023, 0.033, and 0.041 respectively.
}
\end{list}

For the $z\ga 2$ galaxies (see photometric range above, 
Eqs.(\ref{eq:highzselect,1})-(\ref{eq:highzselect,3})):
\setcounter{bean}{0}
\begin{list}%
{(\roman{bean})}{\usecounter{bean}}
\item{$cr$=4: $(B-V)-0.5>(V-I)$: \hskip 5cm (8 galaxies; $z\ga 3$)
\be
\label{eq:Hz2,1}
z_a = 2.37  +0.02\,(U-B) +1.61 \,(B-V) -2.47 \,(V-I),
\hskip 1cm    
\sigma_z= 0.14.
\ee
The 1-$\sigma$ uncertainties of the coefficients are 
0.16, 0.04, 0.23, and 0.40 respectively.
}
\item{$cr$=5: $(B-V)-0.5\leq (V-I)$: \hskip 5cm (6 galaxies; $z\la 3$)
\be
\label{eq:Hz2,2}
z_a =  2.18 +0.10\,(U-B) +0.20 \,(B-V)  + 0.75 \,(V-I),
\hskip 1cm    
\sigma_z= 0.25   
\ee
The 1-$\sigma$ uncertainties of the coefficients are 
0.14, 0.09, 0.34, and 0.77 respectively.
}
\end{list}

We present in Fig.1 the best-fit analytic redshifts $z_a$ (from 
Eqs.(\ref{eq:Lz3,1})-(\ref{eq:Hz2,2})) versus the measured spectroscopic 
galaxy redshifts $z$.
The galaxies with known measurement errors in $UBVI$ are plotted with 
error bars in $z_a$. We used 82 out of 90 galaxies with
available spectroscopic redshifts;
\footnote{We do not count the three $z>2$ galaxies
for which the published spectroscopic redshifts are erroneous or 
very uncertain (M. Sawicki 1998, private communication).}
the eight error bars without points 
in Fig.1 denote the galaxies not used in determining the $z_a$ relations.
These eight outlying galaxies have a mean dispersion of
$\sigma_z\simeq 0.45$ between their estimated redshifts $z_a$ and
their spectroscopic redshifts $z$; they mostly lie near
the boundaries of the five color ranges.
A table listing these eight galaxies is available 
by anonymous ftp in the elt/:HDF subdirectory of astro.princeton.edu.
%
%
%
%
%
%
%
%
%

It is apparent from Fig.1 that the above simple relations provide
good estimates of the galaxy redshifts.
The constant offsets in Eqs.(\ref{eq:Lz3,1})-(\ref{eq:Hz2,2})
represent a rough indicator of the mean galaxy redshift in a given
color range; the larger offsets represent higher redshift galaxies.
The linear color-redshift relation used above is considerably simpler 
than the higher-order polynomial fit used by Connolly et al. (1998),
and has a smaller number of free parameters; it also yields a
smaller dispersion ($\sigma_z \simeq 0.034$ versus 0.097)
for one of the $z\la 2$ color-ranges (see Eq.(\ref{eq:Lz3,1})).

In applying our formulae to the HDF photometric catalog, we first use
Eqs.(\ref{eq:highzselect,1})-(\ref{eq:highzselect,3}) to select
$z\ga 2$ galaxies.
In estimating color redshifts for $z < 2$ galaxies we then use
Eqs.(\ref{eq:Lz3,1})-(\ref{eq:Lz3,3}), and for
$z\ga 2$ redshifts we use Eq.(\ref{eq:Hz2,1})-(\ref{eq:Hz2,2}). 
In the next section, we compare these estimated color
redshifts with those obtained from spectral template fitting.

\section{Comparison With the Template-fitting Method}

Several groups (e.g. \cite{Lanze96}, \cite{GwHa96}, \cite{Saw97})
have obtained photometric redshifts for the HDF galaxies
by comparing the observed $UBVI$ magnitudes with 
a set of galaxy templates of different spectral types 
redshifted to evenly spaced redshifts.
The photometric redshifts obtained by these  
groups are generally consistent with each other, although
large differences exist for some galaxies (\cite{Sawweb97}).

The color-redshift relations derived in the present paper (\S 2)
yield a considerably smaller
dispersion between the estimated and spectroscopic redshifts 
than the template-fitting method;
this occurs because our method explicitly 
minimizes the dispersion for each color range
of galaxies with measured redshifts which are used in fitting
the color-redshift relations.
A comparison of our predicted color redshifts, $z_a$, 
with the photometric redshifts from template-fitting by 
Sawicki et. al. (1997), $z_{temp}$,
is presented in Fig.2 for the 848 HDF galaxies with $I<27$ and with 
measured $UBVI$ magnitudes.
The solid diagonal line in Fig.2 indicates $z_a=z_{temp}$;
the two dotted diagonal lines mark the region in which 
$|z_a-z_{temp}|\leq 0.5$.
The results show that the two estimators, the simple color redshift 
estimator and the template-fitting redshift estimator, are
generally consistent with each other, with some outlyers.
About 90\% of all galaxies lie within $|z_a-z_{temp}|\leq 0.5$.
From the 10\% that lie outside this region (83 out of 848 galaxies),
nearly half lie close to the region's boundary.
The dozen discordant redshifts with $z_{temp} \sim 2$ and 
$z_a <0.5$ (Fig.2) are probably due to the gap in the available HDF 
spectroscopic redshifts at $z \sim 2$; thus galaxies with true redshifts
of $z \sim 2$ may have been assigned wrong redshifts by
the best-fit analytic formulae.
The analytic formulae presented above can be improved
as spectroscopic redshifts are measured for more galaxies in
the HDF (especially in the missing redshift range of $1.4 \leq z \leq 2.2$).

The agreement between the analytic color redshifts we obtain
and the spectral template-fitting photometric redshifts obtained
by Sawicki et al. (1997) is comparable to the
consistency among the various methods utilizing the spectral template-fitting
technique. This illustrates that the linear analytic relations 
based on $UBVI$ colors provide a good and easy method
for estimating galaxy redshifts.
In fact, its dispersion in the different color ranges is lower
than given by the template-fitting methods.

\section{Estimated Redshift Catalog of HDF Galaxies}

We have calculated the estimated redshifts $z_a$ of the HDF galaxies (848 
galaxies with $I<27$ and with measured $UBVI$ nagnitudes)
based on the color relations given above (\S 2). We present
these redshifts in a catalog that is available
by anonymous ftp in the elt/:HDF subdirectory of astro.princeton.edu.

The Estimated Redshift Catalog of HDF Galaxies includes the following
information for each galaxy: galaxy identification number; 
$x$ and $y$ pixel positions on the v2 HDF
image; $UBVI$ magnitudes; our color redshift estimate, $z_a$
(based on Eq.(\ref{eq:Lz3,1})-(\ref{eq:Hz2,2}));
photometric redshift from the template-fitting
method by Sawicki et al, $z_{temp}$;
and, when available, the observed spectroscopic redshift, $z$.
The color range number of each galaxy, $cr$ (\S 2), is also listed 
in the catalog.
Hard copies of the catalog are available upon request.

\section{Redshift Clustering}

We use the estimated color redshifts of the HDF galaxies to investigate 
the large-scale redshift clustering of galaxies.
The small $\sigma_z$ dispersion found for some of
the color ranges provides the possibility of detecting large-scale clustering
among the distant galaxies.

We use the estimated redshifts $z_a$ for the 848 galaxies with
$I< 27$ from the Estimated Redshift Catalog of HDF Galaxies (\S 4). 
We find that most of the galaxies satisfy
either $(U-B)< (B-V) -0.1$ (230 galaxies) or
$(U-B)\geq (B-V) -0.1\leq V-I$ (333 galaxies).
Since the number of galaxies is relatively large and the
dispersion between $z_a$ and $z$ is relatively small
[$\sigma_z=0.034$ and $\sigma_z=0.097$ respectively], 
we use these two groups to study the redshift distribution.

Fig.3 presents the estimated redshift distribution 
[using Eq.(\ref{eq:Lz3,1})] of
230 HDF galaxies with $(U-B)< (B-V) -0.1$ and $I< 27$,
for a bin size of $\Delta z_a=0.03$.
The redshift distribution suggests the existence of peaks
that indicate large-scale clustering of galaxies to $z\sim 1$.
To estimate the significance of the observed structures in the 
redshift distribution, we calculate the expected smoothed distribution
by convolving the observed redshift distribution 
with a Gaussian of width $\sigma_z^r=0.1$.
The dashed and dotted lines in Fig.3 represent the mean smoothed 
distribution and the 1-$\sigma$ contour respectively,
for $\sigma_z^r=0.1$ and $10^4$ realizations of the smoothed distribution.
Most of the observed peaks are marginally significant at
levels of 1 to 3$\sigma$ above the smoothed distribution.
These peaks are consistent with the peaks revealed by spectroscopic 
observations of galaxies to $z \la 0.8$ in this region (\cite{Cohen96});
the location of the spectroscopic peaks are marked by the arrows on top 
of Fig.3. We see that the location of our suggested peaks are consistent
with those seen directly with a smaller number of
spectroscopically measured redshifts.
Fig.3 suggests an additional peak at $z\sim 0.8$ that is not
yet confirmed by spectroscopic data.
The peaks are seen consistently in different parts of the HDF field,
with no evidence of sub-clustering on the sky.
These peaks suggest large-scale clustering in the galaxy distribution at
high redshifts; they may represent the distant ($z\sim 1$) counterpart
to local superclusters, or walls, seen at low redshifts
(\cite{Broad90}; \cite{Bahcall91}).
Most recently, such peaks have also been seen at $z \simeq 3$
(\cite{Stei97}).

Fig.4 presents the redshift distribution of
333 HDF galaxies with $(U-B)\geq (B-V) -0.1\leq V-I$ 
and $I< 27$, for a bin size of $\Delta z_a=0.09$.
(For this group, the redshift dispersion is $\sigma_z=0.097$).
The dashed and the dotted lines in Fig.4 represent the mean smoothed 
distribution and the 1$-\sigma$ contour respectively,
for $\sigma_z^r=0.09$ and $10^3$ realizations of the smoothed distribution.
A peak is suggested at $z_a \sim 1$ and possibly at $z\sim 1.3$, 
but the large dispersion
for this color sub-sample appears to ``wash-out'' any other significant
underlying peaks.


\section{Summary and Discussion}

Using HDF photometric and spectroscopic data,
we have determined a set of simple analytic formulae that yield 
estimated galaxy redshifts to $z\la 4$
in terms of linear combinations of 
three measured colors, 
$U-B$, $B-V$, and $V-I$ (Eqs.(\ref{eq:Lz3,1})-(\ref{eq:Hz2,2})).
The derived analytic formulae in five color ranges exhibit small
dispersions between the estimated and spectroscopic redshifts.
For $z\la 2$ galaxies, the redshift dispersion ranges from
$\sigma_z=0.034$ to $\sigma_z=0.097$ for different color ranges.
For $z\ga 2$ galaxies, we find
$\sigma_z=0.14$ and $\sigma_z=0.36$ for two color ranges
which typically represent $z\ga 3$ and $z\la 3 $ galaxies respectively.
These color-redshift relations apply to about 90
in the sample.

The smallest dispersion between the color and the spectroscopic 
redshifts, $\sigma_z=0.034$, 
occurs for the $z\la 2$ galaxies satisfying $(U-B)< (B-V) -0.1$;
28 galaxies with measured redshifts are used 
in deriving the relation for the estimated redshift,
with only 4 free parameters (the coefficients in Eq.(\ref{eq:Lz3,1})). 
There are 230 HDF galaxies with $I< 27$ and measured UBVI magnitudes
that belong to this color range; we investigate the large-scale
redshift distribution of these galaxies and find evidence for peaks
in the redshift distribution that suggest large-scale clustering
to at least $z\sim 1$.
These results are consistent with those of Cohen et. al. (1996)
using observed spectroscopic redshifts of a smaller number of galaxies.

We have applied our color redshift formulae to the entire HDF photometric 
catalog and find that the derived redshifts are consistent with those
obtained from spectral template-fitting techniques.
The analytic relations, by design, yield lower dispersion
than the template-fitting method.
The color-redshift relations have the advantage of being
simple, model independent, and easy to use. They can be
further improved with additional data.
These analytic color-redshift estimators
are useful in providing empirical estimates
of galaxy redshifts to $z\la 4$ using multiband photometry.

Our Estimated Redshift Catalog of HDF Galaxies, 
based on our color redshift formulae for all 848 HDF galaxies with
$I<27$ and measured $UBVI$ fluxes, is available 
by anonymous ftp in the elt/:HDF subdirectory of astro.princeton.edu.

Note that our color-redshift relations (Eqs.(\ref{eq:Lz3,1})-(\ref{eq:Hz2,2}))
are derived using AB magnitudes
and for the HDF filters. For application to other photometric catalogs,
the appropriate spectroscopic training set should be used; when such a
training set is not available, Eqs.(\ref{eq:Lz3,1})-(\ref{eq:Hz2,2})
may provide useful estimates after appropriate photometric
transformation has been performed between the different filter systems. 
Also note that these color-redshift relations should not be applied
to galaxies which lie close to the boundaries of the color ranges.

Finally, we note that our color-redshift relations are limited by
the absence of measured spectroscopic redshifts for galaxies in the range 
of $1.4 \le z \le 2.2$ (see Fig.1 and Fig.2). 
It is very important to obtain spectroscopic redshifts in this range,
because it will not only enable better calibration of photometric 
redshifts, it will also help us understand the nature of galaxies
in the intermediate redshift range.

\acknowledgements{\centerline{\bf Acknowledgements}}

It is a pleasure to thank Marcin Sawicki for providing
the photometric catalog of the HDF
and the corresponding photometric redshifts,
as well as helpful comments on the manuscript;
Daniela Calzetti for helpful communications concerning spectral templates;
Judy Cohen and David Hogg, who also provided helpful comments on the
manuscript, for the use of redshifts prior to publication;
and James Rhoads for help with IRAF.
This work is partially supported by NSF grants AST93-15368
and AST98-02802.

\clearpage

\figcaption[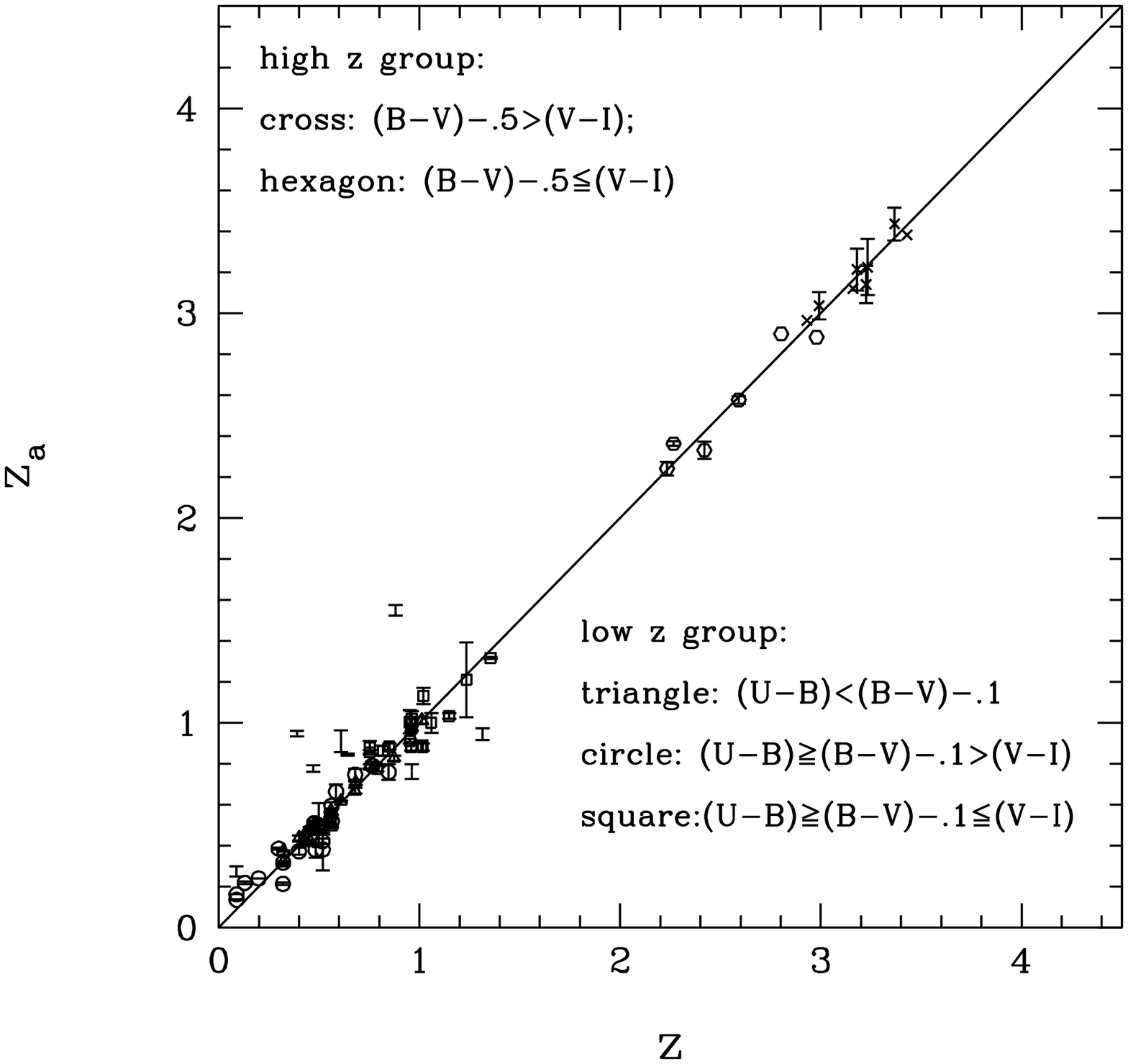]
{Color-based analytic redshift estimate $z_a$ (given by 
Eqs.(\ref{eq:Lz3,1})-(\ref{eq:Hz2,2}))
versus the spectroscopic redshift $z$ for 90 HDF galaxies. 
The galaxies with known measurement errors in $UBVI$ are plotted with error 
bars in $z_a$.}

\figcaption[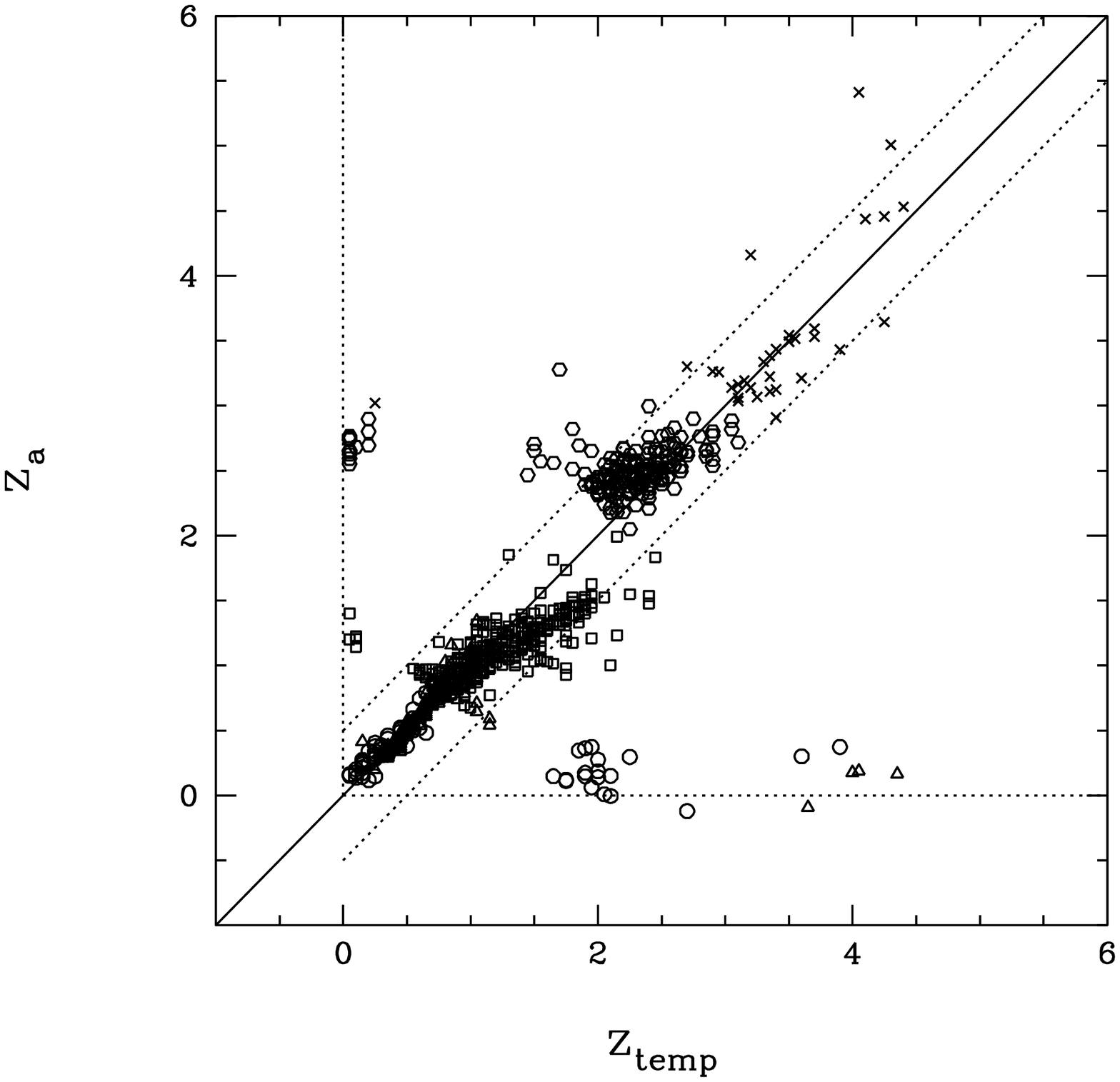]
{Color-based analytic redshift estimate $z_a$ (given by 
Eqs.(\ref{eq:Lz3,1})-(\ref{eq:Hz2,2}))
versus the Sawicki et. al. (1997) template-fitting
photometric redshift $z_{temp}$, for 848 galaxies in the HDF
with $I<27$ and measured $UBVI$. 
The symbols are the same as in Fig.1.
The solid diagonal line indicates $z_a=z_{temp}$;
the dotted lines mark the region $|z_a-z_{temp}|\leq 0.5$.}

\figcaption[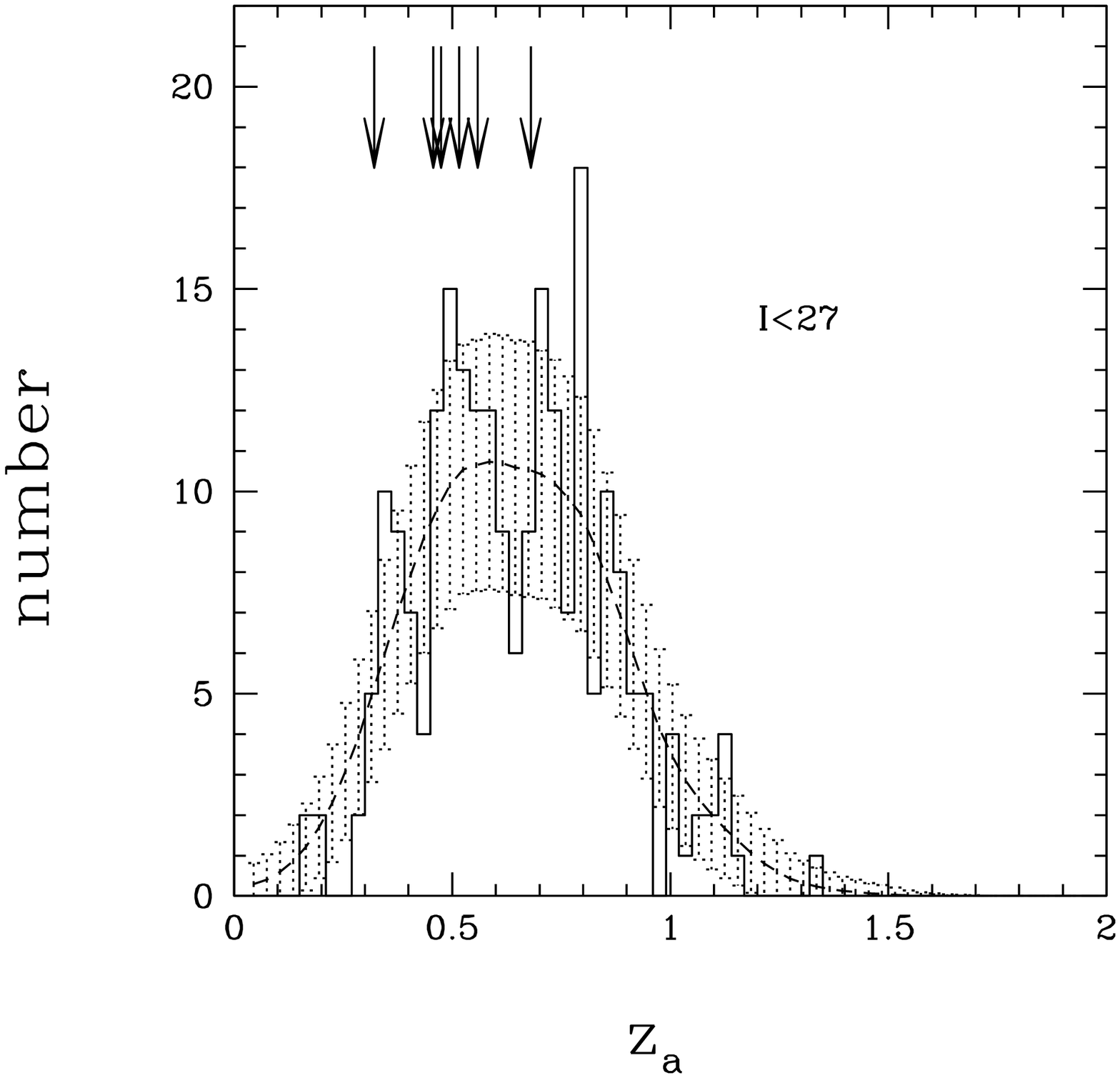]
{The estimated redshift ($z_a$) distribution of
230 galaxies with $(U-B)< (B-V) -0.1$ and $I<27$ in the HDF,
for a bin size of $\Delta z_a=0.03$.
The dashed and the dotted lines represent the mean smoothed 
distribution and the 1$-\sigma$ contour respectively,
for $\sigma_z^r=0.1$ and $10^4$ realizations of the smoothed distribution
(\S 5). The arrows on the top of the figure indicate the location of the
peaks to $z\sim 0.8$ observed from spectroscopic redshifts of
a smaller number of galaxies by Cohen et al. (1996).}

\figcaption[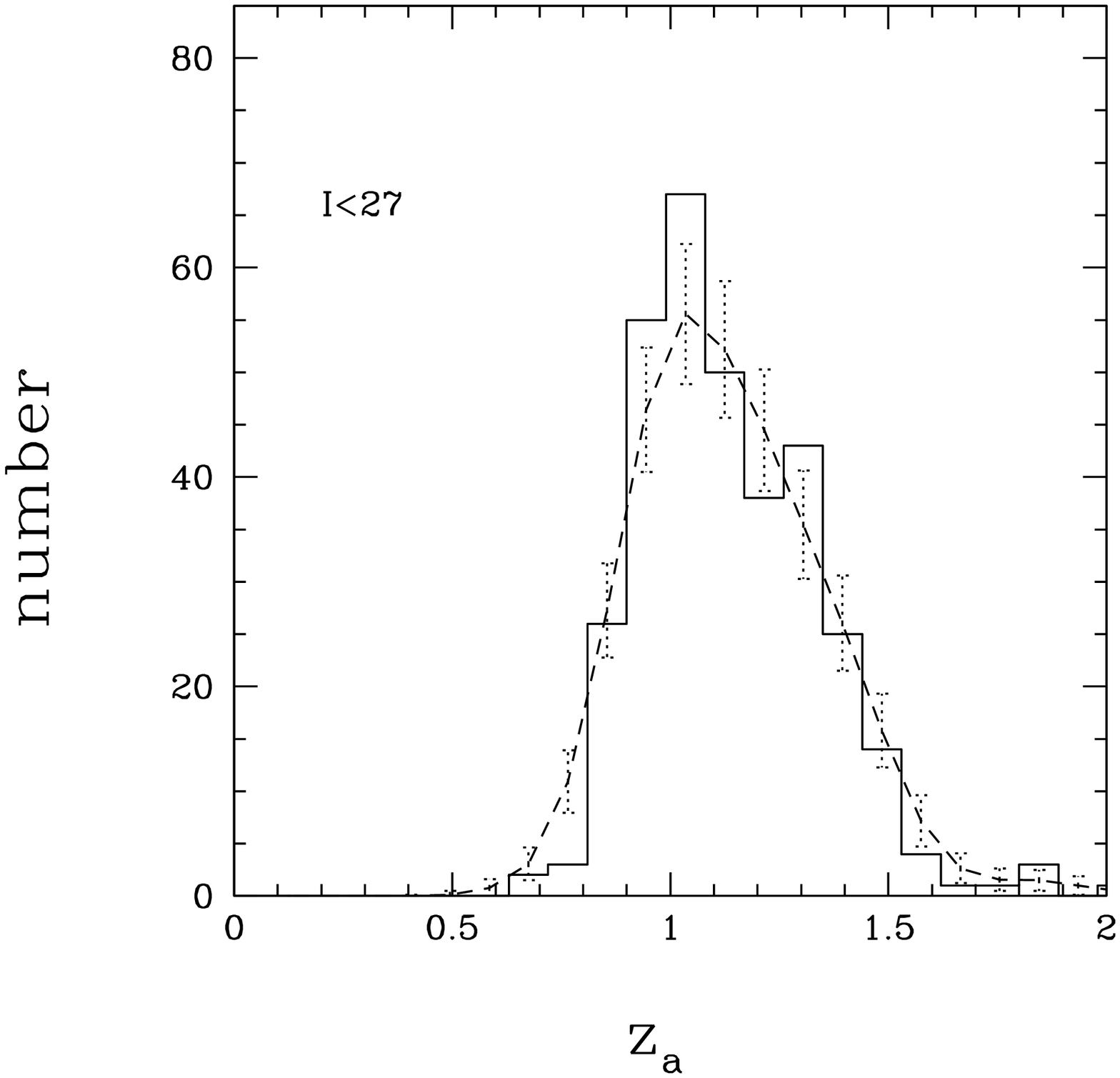]
{The estimated redshift ($z_a$) distribution  of
333 galaxies with $(U-B)\geq (B-V) -0.1\leq V-I$ and $I<27$ 
in the HDF, for a bin size of $\Delta z_a=0.09$.
The dashed and the dotted lines represent the mean smoothed 
distribution and the 1$-\sigma$ contour respectively,
for $\sigma_z^r=0.09$ and $10^3$ realizations of the smoothed distribution
(\S 5).}



%

\clearpage

\setcounter{figure}{0}

\plotone{Wang.fig1.eps}
\figcaption[Wang.fig1.eps]
{Color-based analytic redshift estimate $z_a$ (given by 
Eqs.(\ref{eq:Lz3,1})-(\ref{eq:Hz2,2}))
versus the spectroscopic redshift $z$ for 90 HDF galaxies. 
The galaxies with known measurement errors in $UBVI$ are plotted with error 
bars in $z_a$.}

\plotone{Wang.fig2.eps}
\figcaption[Wang.fig2.eps]
{Color-based analytic redshift estimate $z_a$ (given by 
Eqs.(\ref{eq:Lz3,1})-(\ref{eq:Hz2,2}))
versus the Sawicki et. al. (1997) template-fitting
photometric redshift $z_{temp}$, for 848 galaxies in the HDF
with $I<27$ and measured $UBVI$. 
The symbols are the same as in Fig.1.
The solid diagonal line indicates $z_a=z_{temp}$;
the dotted lines mark the region $|z_a-z_{temp}|\leq 0.5$.}

\plotone{Wang.fig3.eps}
\figcaption[Wang.fig3.eps]
{The estimated redshift ($z_a$) distribution of
230 galaxies with $(U-B)< (B-V) -0.1$ and $I<27$ in the HDF,
for a bin size of $\Delta z_a=0.03$.
The dashed and the dotted lines represent the mean smoothed 
distribution and the 1$-\sigma$ contour respectively,
for $\sigma_z^r=0.1$ and $10^4$ realizations of the smoothed distribution
(\S 5). The arrows on the top of the figure indicate the location of the
peaks to $z\sim 0.8$ observed from spectroscopic redshifts of
a smaller number of galaxies by Cohen et al. (1996).}

\plotone{Wang.fig4.eps}
\figcaption[Wang.fig4.eps]
{The estimated redshift ($z_a$) distribution  of
333 galaxies with $(U-B)\geq (B-V) -0.1\leq V-I$ and $I<27$ 
in the HDF, for a bin size of $\Delta z_a=0.09$.
The dashed and the dotted lines represent the mean smoothed 
distribution and the 1$-\sigma$ contour respectively,
for $\sigma_z^r=0.09$ and $10^3$ realizations of the smoothed distribution
(\S 5).}

\end{document}